\documentclass[11pt]{article}
\usepackage{amsfonts}
\usepackage{amsmath}
\begin{document}
\begin{titlepage}
\begin{flushright}
CP3-06-07\\
ICMPA-MPA/2006/34\\
August 2006\\
\end{flushright}
\begin{centering}
 
{\ }\vspace{1cm}
 
{\Large\bf Topological Background Fields as}

\vspace{5pt}

{\Large\bf Quantum Degrees of Freedom of Compactified Strings}\\

\vspace{1.5cm}

Jan Govaerts$^{\rm 1,2}$ and Florian Payen$^{\rm 1}$

\vspace{1.3cm}

$^{\rm 1}${\em Center for Particle Physics and Phenomenology (CP3)}\\
{\em Institute of Nuclear Physics}\\
{\em Department of Physics, Catholic University of Louvain}\\
{\em 2, Chemin du Cyclotron, B-1348 Louvain-la-Neuve, Belgium}\\
{\em E-mail: {\tt Jan.Govaerts@fynu.ucl.ac.be, Payen@fynu.ucl.ac.be}}

\vspace{1.0cm}

$^{\rm 2}${\em International Chair in Mathematical Physics and 
Applications (ICMPA)}\\
{\em University of Abomey-Calavi}\\
{\em 072 B.P. 50, Cotonou, Republic of Benin}\\

\vspace{1.0cm}

\begin{abstract}
\noindent
It is shown that background fields of a topological character usually
introduced as such in compactified string theories correspond to quantum
degrees of freedom which parametrise the freedom in choosing a
representation of the zero mode quantum algebra in the presence
of non-trivial topology. One consequence would appear to be that
the values of such quantum degrees of freedom, in other words
of the associated topological background fields, cannot be determined
by the nonperturbative string dynamics.

\end{abstract}

\vspace{10pt}

\end{centering} 

\vspace{25pt}

\end{titlepage}

\setcounter{footnote}{0}

\section{\bf Introduction}
\label{Sect1}

The study of string theories compactified onto
a large variety of spaces with different topologies and geometries
has produced profound insights into the nonperturbative properties 
of these systems, ultimately leading to the discovery of D-branes 
and the web of dualities relating all known superstring 
theories to the yet to be constructed underlying M-theory (for reviews
and references to the original literature, see Refs.~\cite{Joe,Cliff}).
Even in the simplest framework of flat torus compactification, besides 
the geometry of the internal manifold, further background fields
of a topological character such as an antisymmetric tensor and
Wilson lines are introduced.\cite{Ginsparg} Analogous background fields also
exist for more intricate compactifications with richer topology and geometry. 
Even though one could invoke possible dynamical mechanisms that
would lead to non-vanishing expectation values of such background fields given 
a specific compactification topology, it may seem still somewhat unsettling 
to have to introduce in some ad-hoc fashion such background fields 
into a theory which purportedly ought to define the ultimate 
fundamental framework for all of matter and its quantum interactions.

In this brief note, we wish to point out that in the presence of
compactifications of non-trivial topology possessing non-contractible
cycles, namely possessing a non-trivial fundamental or first homotopy group, 
there exist specific quantum degrees of freedom to which classical 
string theory is oblivious. These quantum degrees of freedom play 
precisely a role akin to that of topological background fields,
independently from the metric tensor specifying the compactified geometry 
of which the value is presumably determined dynamically. The values 
of these quantum degrees of freedom are related to a choice
of representation of the Heisenberg algebra for the zero mode
degrees of freedom of the string vibrating in the compactified
dimensions. From that point of view, the situation is somewhat similar
to that with spin for a rotationally invariant system: which representation
of the rotation group algebra is to be used for the quantised system
is matter of the experimental determination of the spin of the
physical system under consideration. There is no known dynamical
framework which would predict the spin value, say, of the electron.
Likewise, values for the topological background fields in string
compactifications besides the compactified geometry
may possibly not be set through dynamical considerations, 
but could remain contingent on a choice of representation
for specific algebraic structures intrinsic to quantum dynamics
in the presence of non-trivial topology.

We shall not present here a detailed discussion for large
classes of superstring compactifications, but restrict to
the core facts we wish to emphasize by staying within the
simplest context of the toroidal compactification of free 
oriented bosonic strings in Minkowski spacetime. As is well known,
from a quantum point of view the oscillation modes of the 
string correspond to creation and annihilation operators, 
whereas the position and momentum operators describing the motion 
of its centre of mass satisfy the Heisenberg algebra. 
However, as discussed once again in a recent paper\cite{Gov}
even though such results have been known for a long time\cite{Morette,Schulmann}
but not as widely as they would deserve to be, contrary to the situation in the case of
Euclidean space, the Heisenberg algebra admits an infinity
of unitarily inequivalent representations if the position operator takes 
its eigenvalues on a manifold with a non-trivial first homotopy
group. This is the case for the string centre of mass 
if some spatial dimensions are compactified on a torus, for
instance. The purpose of this note is to account for 
the existence of these inequivalent representations of the
Heisenberg algebra, namely genuine quantum degrees of freedom
which decouple from the classical dynamics, in torus 
compactifications of oriented bosonic strings.\cite{Payen1}
Extensions to other classes of superstring theories and/or
compactifications should also be of interest.

Our presentation is organised as follows. First the quantum description 
of the motion of a pointlike object on an arbitrary manifold, 
and the construction of the inequivalent representations of 
the corresponding Heisenberg algebra, as described in Ref.~\cite{Gov},
are recalled in the next section. Then Sec.~\ref{Sec3} particularises 
the discussion to the case of a Minkowski spacetime with some spatial
dimensions compactified onto a torus. Next in Sec.~\ref{Sec4} the analysis
is applied specifically to the well known canonical quantisation of open and closed
oriented bosonic strings on this compactified spacetime (see, {\it e.g.},
Refs.~\cite{Joe,Cliff,GSW,LSW,Rabi}). Finally Sec.~\ref{Sec5} addresses
some possible implications of our results.

\section{The Heisenberg Algebra}
\label{Sec2}

Let us first consider a physical system describing the motion of a
pointlike object\footnote{The possible generalisation of similar
considerations to extended objects is an open issue.} on an arbitrary 
connected oriented manifold $\mathcal{M}$ of dimension ${\cal N}$. 
Units such that $\hbar=1=c$ are used throughout.

Classically, within the Hamiltonian setting, the state of
the system is described by local configuration
space coordinates $q^{n}$ and their associated canonical
conjugate momenta $p_{n}$, with the Poisson brackets
\begin{equation}
\{q^{n},q^{m}\} =  0, \qquad 
\{p_{n},p_{m}\} =  0, \qquad 
\{q^{n},p_{m}\} =  \delta^{n}_{m}, \qquad n,m=1,2,\cdots,{\cal N}.
\end{equation}
Following the usual rules of canonical quantisation, one associates
to these classical degrees of freedom linear self-adjoint
operators $\hat{q}^{n}$ and $\hat{p}_{n}$ acting on the ``Hilbert"
space of quantum states, satisfying the Heisenberg algebra
\begin{equation}
[\hat{q}^{n},\hat{q}^{m}] = 0, \qquad 
[\hat{p}_{n},\hat{p}_{m}] = 0, \qquad 
[\hat{q}^{n},\hat{p}_{m}] = i\delta^{n}_{m}, \qquad n,m=1,2,\cdots,{\cal N}.
\label{eq:Heis}
\end{equation}

In order to represent the Heisenberg algebra on the Hilbert space,
let us assume there exists a basis of eigenvectors $|q\rangle$ of
the configuration space or position operators $\hat{q}^{n}$,
\begin{equation}
\hat{q}^{n}|q\rangle=q^{n}|q\rangle,
\end{equation}
in one-to-one correspondence with the points of the manifold
$\mathcal{M}$. Assuming implicitly a positive definite
hermitian inner product on the space of quantum states,
the normalisation of the position eigenstates,
\begin{equation}\label{g3}
\langle q|q' \rangle=\frac{1}{\sqrt{g(q)}}\delta(q-q'),
\end{equation}
may always be expressed in terms of a volume ${\cal N}$-form
\begin{equation}
\Omega(q)=\sqrt{g(q)}\,dq^{1}\wedge\ldots\wedge dq^{{\cal N}}.
\end{equation}
Position operators act on a quantum state $|\psi\rangle$ according to
\begin{equation}
\langle q|\hat{q}^{n}|\psi\rangle=q^{n}\langle q|\psi\rangle,
\end{equation}
while it may be shown, based on the abstract Heisenberg algebra
in (\ref{eq:Heis}), that the action of the momentum operators
may be parametrised according to
\begin{equation}
\langle q|\hat{p}_{n}|\psi\rangle=
\frac{-i}{g^{1/4}(q)}\left[\frac{\partial}{\partial q^{n}}+i
A_{n}(q)\right]g^{1/4}(q)\langle q|\psi\rangle
\end{equation}
in terms of a real closed 1-form
\begin{equation}
A(q)=A_{n}(q)dq^{n},\qquad dA(q)=0.
\end{equation}
The real forms $\Omega(q)$ and $A(q)$ characterise the considered
representation of the Heisenberg algebra.

Under a general change of position eigenstate basis in Hilbert space
of the following form,
\begin{equation}
|q\rangle^{(2)}=R(q)\,e^{i\theta(q)}\,|q\rangle^{(1)},
\end{equation}
namely through a simple change of phase and normalisation
both left unspecified by the above considerations, and
which thus relates, up to their normalisation, unitarily
equivalent representations\footnote{Strictly speaking, a change of
normalisation does not define a unitary transformation. However since
the normalisation of position eigenstates, parametrised through the
volume form, is of no physical relevance, only local phase redefinitions
of these eigenstates associated to any 0-form $\theta(q)$ on ${\cal M}$ are
considered to define unitary equivalent realisations of the Heisenberg
algebra.} of the Heisenberg algebra, the two relevant forms characteristic
of any such representation transform as
\begin{equation}
\Omega^{(2)}(q)=\frac{1}{R^2(q)}\,\Omega^{(1)}(q), \qquad
A^{(2)}(q)=A^{(1)}(q)+d\theta(q).
\end{equation}
Hence unitarily inequivalent representations of the Heisenberg
algebra are in one-to-one correspondence with the equivalence 
classes of closed 1-forms differing by exact 1-forms. 
These classes constitute the 1-cohomology of the manifold $\mathcal{M}$, 
which is non-trivial if the manifold is not simply connected.
In the case of the simply connected Euclidean configuration space manifold,
one recovers the well known Stone--von Neumann theorem stating that
for Euclidean spaces, up to arbitrary unitary transformations
there exists a single representation of the Heisenberg algebra
given by the usual plane wave realisations.

\section{The Heisenberg Algebra and Torus Compactification}
\label{Sec3}

Let us now particularise our discussion to the motion of a
pointlike object on a Minkowski spacetime of total dimension
$D=D_1+D_2$, where $D_{2}$ spatial dimensions are compactified
onto a Euclidean torus.

To represent this direct product compactified spacetime, first,
one has a Minkowski spacetime $\mathrm{M}^{D_1}$ of dimension $D_1$
with coordinates $x^{\mu}$ and metric
\begin{equation}
\eta_{\mu\nu}=diag(-1,1,\ldots,1),\qquad
\mu,\nu=0,1,\cdots,D_1-1,
\end{equation}
and second, one has a Euclidean space $\mathrm{E}^{D_2}$ of
dimension $D_{2}$, with cartesian coordinates $y^{I}$ and 
Euclidean metric
\begin{equation}
\delta_{IJ}=diag(1,\ldots,1), \qquad I,J=1,2,\cdots,D_2.
\end{equation}
On the space $\mathrm{E}^{D_2}$ and its dual $\mathrm{E}^{D_2*}$,
let us then introduce dual bases $\{{e_{a}}^{I}\}$ and
$\{e^{*a}_{I}\}$, with $e^{*a}_{I}{e_{b}}^{I}=\delta^{a}_{b}$,
generating the lattices
\begin{equation}
\mathrm{\Lambda}=\big\{l^{I}=l^{a}{e_{a}}^{I}:
l^{a}\in\mathbb{Z}\big\}, \qquad
\mathrm{\Lambda}^{*}=\big\{k_{I}=k_{a}e_{I}^{*a}:
k_{a}\in\mathbb{Z}\big\}.
\end{equation}
The torus $\mathrm{T}^{D_2}$ is then constructed by identifying points
of $\mathrm{E}^{D_2}$ of the form $y^{I}+2\pi l^{I}$, with $l^{I}$
in $\mathrm{\Lambda}$. Finally the $D$-dimensional
compactified spacetime is the direct product
$\mathrm{M}^{D_1}\times\mathrm{T}^{D_2}$.

The compactified spacetime
$\mathrm{M}^{D_1}\times\mathrm{T}^{D_2}$ is not simply connected,
and its 1-cohomology is thus non-trivial. Consequently the Heisenberg
algebra of the position, $\hat{x}^{\mu},\hat{y}^{I}$, and momentum,
$\hat{p}_{x\mu},\hat{p}_{yI}$, operators admits inequivalent
representations on the space of quantum states.
Each of these representations may univocally be characterised,
on the one hand, say by the canonical unit $D$-form associated
to the Minkowskian geometry
\begin{equation}
\Omega=dx^{1}\wedge\ldots\wedge dx^{D_1}\wedge
dy^{1}\wedge\ldots\wedge dy^{D_2},
\end{equation}
and on the other hand, by a real closed 1-form
\begin{equation}
A=A_I\,dy^I
\end{equation}
associated to a constant vector $A_{I}$ in the unit cell of the
dual lattice $\mathrm{\Lambda}^{*}$. Introducing a basis of
eigenvectors $|x,y\rangle$ of the position operators
$\hat{x}^{\mu},\hat{y}^{I}$, with the normalisation
\begin{equation}
\langle x,y| x',y' \rangle = \delta(x-x')\,\delta(y-y'),
\end{equation}
the action of the position and momentum operators on
a quantum state $|\psi\rangle$ may be expressed as
\begin{equation}
\langle x,y | \hat{x}^{\mu} | \psi \rangle=x^{\mu}\langle x,y | \psi \rangle, 
\qquad \langle x,y | \hat{p}_{x\mu} | \psi
\rangle=-i\frac{\partial}{\partial x^{\mu}}\langle x,y | \psi \rangle,
\end{equation}
and
\begin{equation}
\langle x,y | \hat{y}^{I} | \psi \rangle=y^{I}\langle x,y | \psi
\rangle, \qquad \langle x,y | \hat{p}_{yI} | \psi
\rangle=-i\big(\frac{\partial}{\partial y^{I}}+iA_{I}\big)\langle
x,y | \psi \rangle.
\end{equation}

Finally it is useful to identify in the Hilbert space a basis of vectors
$|p_x,p_y\rangle$ diagonalising the momentum operators
$\hat{p}_{x\mu},\hat{p}_{yI}$. Their configuration space
wave functions obey the differential equations
\begin{equation}
-i\frac{\partial}{\partial x^{\mu}}\langle x,y|p_x,p_y \rangle
=p_{x\mu}\langle x,y|p_x,p_y \rangle
\end{equation}
and
\begin{equation}
-i\big(\frac{\partial}{\partial y^{I}}+iA_{I}\big)\langle
x,y|p_x,p_y \rangle=p_{yI}\langle x,y|p_x,p_y \rangle,
\end{equation}
with the general solution
\begin{equation}
\langle x,y|p_x,p_y \rangle=C\,\exp i\big(p_x\cdot x+(p_y-A)\cdot y\big),
\end{equation}
$C$ being some arbitrary integration constant.
The operators $\hat{p}_{x\mu}$ admit real eigenvalues
$p_{x\mu}=p_{\mu}$, while single-valuedness or periodicity
conditions on the torus
\begin{equation}
\langle x,y+2\pi l|p_x,p_y \rangle=\langle x,y|p_x,p_y \rangle,
\qquad l^I\in\Lambda,
\end{equation}
restrict the eigenvalues of the operators $\hat{p}_{yI}$ to the
quantised spectrum $p_{yI}=k_{I}+A_{I}$, where $k_{I}$ is a vector
of the dual lattice $\mathrm{\Lambda}^{*}$. Imposing the
normalisation conditions
\begin{equation}
\langle p,k+A|p',k'+A\rangle = \delta(p-p')\,\delta_{kk'}
\end{equation}
and fixing the arbitrary phase to unity, the configuration
space wave functions read
\begin{equation}
\langle x,y | p,k+A\rangle = (2\pi)^{-D/2}V^{-1/2} \, 
\exp i(p\cdot x+k\cdot y),
\end{equation}
where $V$ is the volume of the torus.

\section{Torus Compactification of Bosonic Strings}
\label{Sec4}

We are now ready to address open and closed oriented bosonic strings on the
toroidally compactified spacetime, focusing on the motion of the centre
of mass in light of the previous discussion.

Consider a free relativistic oriented bosonic string. Its world-sheet,
parametrised by coordinates $\sigma^\alpha$ ($\alpha=0,1$) with
$\sigma^0\equiv\tau$ in $[\tau_1,\tau_2]$ and
$\sigma^1\equiv\sigma$ in $[0,\pi]$, is equipped with an intrinsic
metric $\gamma_{\alpha\beta}(\tau,\sigma)$ of signature $(-,+)$,
and embedded through functions $x^{\mu}(\tau,\sigma)$ and
$y^{I}(\tau,\sigma)$ into the compactified spacetime
$\mathrm{M}^{D_1}\times\mathrm{T}^{D_2}$, with total dimension $D$
restricted to the critical value, $D=26$, for the usual reasons
of quantum consistency. The dynamics of the system is determined by 
the linear Polyakov action
\begin{equation}
S[x,y,\gamma]=-\frac{1}{4\pi\alpha'}\int_{\tau_{1}}^{\tau_{2}}\ d\tau
\int_{0}^{\pi}\ d\sigma\ \sqrt{-\det\gamma}\ \gamma^{\alpha\beta}
\left[\eta_{\mu\nu}\partial_{\alpha}x^{\mu}\partial_{\beta}x^{\nu}
+\delta_{IJ}\partial_{\alpha}y^{I}\partial_{\beta}y^{J}\right],
\end{equation}
where $\alpha'$ is the Regge slope.

The usual Hamiltonian analysis and canonical quantisation of 
the system is readily achie\-ved. Local gauge symmetries associated 
to world-sheet diffeomorphisms and Weyl transformations are partly
fixed by choosing the conformal gauge. Results remain identical 
of course when working in the light-cone gauge.\footnote{The latter has
to be properly defined in the presence of compactified spacelike
dimensions, namely the gauge fixed light-cone coordinate must not
involve any of the compactified components.}

\subsection{Open strings}
\label{Sec4.1}

Let us first concentrate on the case of open oriented strings. In the
conformal gauge, the equations of motion
\begin{equation}
\partial^2_\tau x^\mu-\partial^2_\sigma x^\mu=0,\qquad \partial^2_\tau
y^I-\partial^2_\sigma y^I=0,
\end{equation}
together with the Neumann boundary conditions
\begin{equation}
\partial_\sigma x^\mu(\tau,0)=\partial_\sigma
x^\mu(\tau,\pi)=0,\qquad \partial_\sigma
y^I(\tau,0)=\partial_\sigma y^I(\tau,\pi)=0,
\end{equation}
lead to the classical solutions
\begin{eqnarray}
x^{\mu}(\tau,\sigma) & = & x^{\mu}+2\alpha'p_{x}^{\mu}\tau
+i\sqrt{2\alpha'}\sum_{n\in\mathbb{Z}_{0}}\frac{1}{n}
\alpha_{n}^{\mu}e^{-in\tau}\cos n\sigma,\\
y^{I}(\tau,\sigma) & = & y^{I}+2\alpha'p_{y}^{I}\tau
+i\sqrt{2\alpha'}\sum_{n\in\mathbb{Z}_{0}}\frac{1}{n}
\alpha_{n}^{I}e^{-in\tau}\cos n\sigma.
\end{eqnarray}
At the quantum level, the zero modes $\hat{x}^{\mu},\hat{y}^{I}$
and $\hat{p}_{x\mu},\hat{p}_{yI}$ satisfy a Heisenberg algebra,
while the oscillator modes
$\hat{\alpha}_{m}^{\mu},\hat{\alpha}_{m}^{I}$ correspond to
creation or annihilation operators according to whether the integer mode
index $m$ is negative or positive, respectively. The space of quantum
states is obtained as the tensor product of Heisenberg and Fock
representation spaces.

The freedom available in the choice of Heisenberg representation for
the compactified zero modes and parametrised by a vector $A_I$
in the fundamental cell of the dual lattice $\Lambda^*$,
implies that one could consider different types or sectors of such open strings,
each characterised by a different value of $A_I(\alpha)$
distinguished by the label $\alpha$. This situation
is reminiscent of that for spinning strings for which,
depending on the boundary conditions for the world-sheet
spinors, one obtains Ramond or Neveu--Schwarz sectors distinguished
by the representations of the fermionic Fock algebra, and in particular
that of its zero mode sector. 

More generally still, by adding Chan--Paton factors to the ends
of open strings, we seem to be free to choose 
independently in each sector $(i,j)$ ($1\leq i,j \leq N$), possibly
further distinguished by the label $\alpha$, an arbitrary representation of 
the Heisenberg algebra, characterised by a vector $A^{ij}_{I}(\alpha)$ 
in the unit cell of the dual lattice $\mathrm{\Lambda}^{*}$. 

However a consistent description of string interactions requires 
charge and momentum conservation. These conservation laws lead, 
for the annihilation of two, three or more strings, to relations such as
\begin{equation}
A^{ij}_{I}(\alpha)+A^{ji}_{I}(\beta)=0\ ({\rm mod}\,\Lambda^*),\qquad 
A^{ij}_{I}(\alpha)+A^{jk}_{I}(\beta)+A^{ki}_{I}(\gamma)
=0\ ({\rm mod}\,\Lambda^*),
\qquad \ldots
\label{eq:Condi1}
\end{equation}
where the values of the sector labels $\alpha$, $\beta$ and $\gamma$
must be correlated in a fashion to be specified, in relation to
the properties of the string interactions being considered. 

A general analysis of the consequences of the freedom, of a topological
origin, in the choice of quantum degrees of freedom $A^{ij}_I(\alpha)$
for the compactified bosonic zero mode algebra would be of interest.
In this note we restrict solely to the simpler setting, namely that
in which there is only a single type of open string in each Chan--Paton
sector, the label $\alpha$ above distinguishing the choices $A^{ij}_I(\alpha)$
then taking a single value. In that case, the solutions for the vectors
$A^{ij}_{I}$ are of the form
\begin{equation}
A^{ij}_{I}=A^{i}_{I}-A^{j}_{I},\quad i,j=1,2,\cdots,N,
\end{equation}
with $A^i_I$ in the fundamental cell of the dual lattice $\Lambda^*$.

Physical states are defined by the usual annihilation Viraroso constraints.
In particular, the zero mode generator inclusive of the quantum subtraction
constant required by conformal invariance,
\begin{equation}
\hat{L}_{0}=\alpha'\hat{p}_{x}^{2}+\alpha'\hat{p}_{y}^{2}+ \hat{N}-1,
\end{equation}
where $\hat{N}$ is the total string level excitation operator, leads to 
the following mass spectrum
\begin{equation} \label{a1}
\alpha'M^{2}=\alpha'\big(k+A^{i}-A^{j}\big)^{2}+N-1,
\end{equation}
where $k_{I}$ is a vector of the dual lattice
$\mathrm{\Lambda}^{*}$, and $N$ a positive integer specifying the
$\hat{N}$ eigenvalue.

\subsection{Closed strings}
\label{Sec4.2}

Let us now turn to the case of closed oriented strings. In the conformal
gauge, the equations of motion
\begin{equation}
\partial^2_\tau x^\mu-\partial^2_\sigma x^\mu=0,\qquad \partial^2_\tau
y^I-\partial^2_\sigma y^I=0,
\end{equation}
together with the periodicity conditions
\begin{equation}
x^\mu(\tau,\pi)=x^\mu(\tau,0),\qquad
y^I(\tau,\pi)=y^I(\tau,0)+2\pi l^I,
\end{equation}
where the vector $l^{I}$ in the lattice $\mathrm{\Lambda}$
parametrises the torus winding sector of the string configuration,
possess the following classical solutions
\begin{eqnarray}
x^{\mu}(\tau,\sigma) & = &
x^{\mu}+2\alpha'p_{x}^{\mu}\tau+\frac{i}{2}\sqrt{2\alpha'}
\sum_{n\in\mathbb{Z}_{0}}\frac{1}{n}
\left[\alpha_{n}^{\mu}\,e^{-2in(\tau-\sigma)}
+\bar{\alpha}_{n}^{\mu}\,e^{-2in(\tau+\sigma)}\right],\\
y^{I}(\tau,\sigma) & = & y^{I}+2\alpha'p_{y}^{I}\tau+2l^{I}\sigma+
\frac{i}{2}\sqrt{2\alpha'}\sum_{n\in\mathbb{Z}_{0}}\frac{1}{n}
\left[\alpha_{n}^{I}\,e^{-2in(\tau-\sigma)}
+\bar{\alpha}_{n}^{I}\,e^{-2in(\tau+\sigma)}\right].
\end{eqnarray}
At the quantum level, in a given winding sector $l^I$, the zero modes
$\hat{x}^{\mu},\hat{y}^{I}$ and $\hat{p}_{x\mu},\hat{p}_{yI}$
satisfy a Heisenberg algebra, while the oscillator modes
$\hat{\alpha}_{m}^{\mu},\hat{\alpha}_{m}^{I}$ and
$\hat{\bar{\alpha}}_{m}^{\mu},\hat{\bar{\alpha}}_{m}^{I}$
correspond to creation or annihilation operators. The
space of quantum states is constructed as the tensor product of
Heisenberg and Fock representation spaces.

When accounting for all possible winding sectors of the string, one
may apparently freely choose independently in each sector $l^{I}$
an arbitrary representation of the Heisenberg algebra
characterised by a vector $A_{I}(l)$ in the unit cell of the dual
lattice $\mathrm{\Lambda}^{*}$.\footnote{Here again we ignore the
possibility of still more generality, which would be afforded by
including different string sectors $A_I(l;\alpha)$ in the manner
discussed previously for the open string case.} However once again winding
and momentum conservation laws lead, for the annihilation of two, three or more
strings, to restrictions such as
\begin{eqnarray}
&&A_{I}(l_1)+A_{I}(l_2)=0\ ({\rm mod}\,\Lambda^*),\qquad\qquad\qquad l_1+l_2=0, \nonumber\\ 
&&A_{I}(l_1)+A_{I}(l_2)+A_{I}(l_3)=0\ ({\rm mod}\,\Lambda^*),\qquad  l_1+l_2+l_3=0,
\qquad \ldots
\end{eqnarray}
which suggest a general solution of the linear form
\begin{equation}
A_{I}(l)=B_{IJ}\,l^{J},
\end{equation}
where the real coefficients $B_{IJ}$ may be viewed as defining some
2-index covariant tensor of the compactified Euclidean space.
However, since $A_I(l)$ is defined modulo a vector of the dual
lattice $\Lambda^*$, $B_{IJ}$ itself is defined modulo a 2-tensor 
of $\mathrm{\Lambda}^{*}$, namely $B_{IJ}$
belongs only to the ``fundamental cell" of such 2-tensors.
Furthermore $B_{IJ}$ may be decomposed into a symmetric
and an antisymmetric part in the two indices $I$ and $J$.

Among the Virasoro constraints defining physical states, the zero mode
generators
\begin{equation}
\hat{L}_{0}=\frac{\alpha'}{4}\hat{p}_{x}^{2}
+\frac{\alpha'}{4}\big(\hat{p}_{y}-\alpha'{}^{-1}\hat{l}\,\big)^{2}+
\hat{N}-1, \qquad
\hat{\bar{L}}_{0}=\frac{\alpha'}{4}\hat{p}_{x}^{2}
+\frac{\alpha'}{4}\big(\hat{p}_{y}+\alpha'{}^{-1}\hat{l}\,\big)^{2}+
\hat{\bar{N}}-1,
\end{equation}
with $\hat{N}$ and $\hat{\bar{N}}$ the total string level excitation 
operators for each of the world-sheet chiral sectors,
lead to the following mass spectrum
\begin{equation}\label{a2}
\alpha'M^{2}=\alpha'\big(k+Bl\big)^{2}+\alpha'^{-1}l^{2}+2N+2\bar{N}-4
\end{equation}
as well as the level matching condition
\begin{equation}
N-\bar{N}=k_I l^I+B_{IJ}l^{I}l^{J},
\end{equation}
where $l^{I}$ is a winding vector of the lattice $\mathrm{\Lambda}$,
$k_{I}$ a momentum vector of the dual lattice $\mathrm{\Lambda}^{*}$, and
$N$, $\bar{N}$ positive integers. 

In order that whatever the winding sector the set of physical
states be non-empty, we must further require that the
symmetric part of $B_{IJ}$ is a 2-tensor of the 
dual lattice $\Lambda^*$, while since $A_I(l)$ is defined up
to vectors in the dual lattice anyway, one may simply
restrict $B_{IJ}$ to be purely antisymmetric and thus be lying
in the fundamental cell of antisymmetric 2-tensors of
$\Lambda^*$. Consequently, the level matching condition finally reads
\begin{equation}\label{a3}
N-\bar{N}=k_I l^I.
\end{equation}

Furthermore, it may easily be shown that such a choice of representation
of the Heisenberg algebra, characterised by an antisymmetric tensor
$B_{IJ}$ defined modulo the dual lattice $\Lambda^*$, is also
consistent with one-loop modular invariance of the partition
function of the closed string. Given that our results coincide, as indicated
hereafter, with those of closed string torus compactification
in presence of a constant background antisymmetric field $B_{IJ}$ for which
modular invariance is well established,\cite{Ginsparg}
the same conclusion readily follows in our setting. Any other form chosen for $A_I(l)$ 
does not seem capable of meeting the requirement of modular invariance
necessary for the quantum consistency of interacting strings.\cite{Payen1}

\section{Discussion and Perspectives}
\label{Sec5}

In this note, general results relevant to the Heisenberg algebra 
on an arbitrary manifold of non-trivial topology have been applied
to the canonical quantisation of open and closed oriented bosonic 
strings on a toroidally compactified Minkowski spacetime.
Taking into account inequivalent representations of the
algebra describing the centre of mass of the strings, which then
arise from the non-trivial spacetime topology, results of interest
have been established, in particular through the physical mass spectra
of such string models.

In fact, these spectra are well known in the
literature,\cite{Joe,Cliff,Ginsparg,GSW,LSW,Rabi}
but they are then obtained by quantising
strings interacting with constant external topological fields
in the compactified spatial dimensions (whether an antisymmetric
tensor or Wilson lines) in addition to the metric background, while choosing
the trivial representation for the algebra describing their centre of mass.
As a matter of fact, the open string spectrum (\ref{a1}) can be reproduced by
coupling the ends of an open string to a constant diagonal $U(N)$
gauge field\footnote{
In the literature, the spectrum of an open string in a constant
diagonal $U(N)$ gauge field is determined\cite{Joe,Cliff} by analogy with
that of a particle in a constant $U(1)$ gauge field. In fact, this spectrum
follows from the action (\ref{act}), where $K$ is the highest
weight of the fundamental representation of $U(N)$, and $g_0(\tau)$ and
$g_\pi(\tau)$ in $U(N)$ describe the Chan--Paton degrees of freedom
attached to both ends of the string. See Ref.~\cite{Payen2} for a detailed
discussion.}
$A_I=diag(A_I^1,\ldots,A_I^N)$
\begin{equation}\label{act}
S_A[x,y,\gamma, g_0, g_\pi] =
S[x,y,\gamma]+i\int_{\tau_{1}}^{\tau_{2}}\!\!\!d\tau\
Tr\big[K g^{-1}(\partial_{\tau}+iA_I\partial_{\tau}y^I)g\big]
\Big|^{\sigma=\pi}_{\sigma=0}.
\end{equation}
Under T-duality to the D-brane picture, the values for the parameters $A^i_I$ 
are known to correspond to positions of the D-branes on the compactified torus.\cite{Joe,Cliff}
In the present setting, these positions are seen to correspond,
through T-duality, to quantum degrees of freedom arising from the non-trivial
representation theory of the bosonic zero mode Heisenberg algebra
in the presence of non-trivial topology. In the same way, the closed string
spectrum (\ref{a2}), together with the level matching condition (\ref{a3}),
can be reproduced by coupling a closed string to a constant antisymmetric
tensor field $B_{IJ}$\cite{Ginsparg}
\begin{equation}
S_B[x,y,\gamma]=S[x,y,\gamma]
-\frac{1}{4\pi}\int_{\tau_{1}}^{\tau_{2}}\!\!\!d\tau\int_{0}^{\pi}\!\!\!
d\sigma\
\epsilon^{\alpha\beta}B_{IJ}\partial_{\alpha}y^{I}\partial_{\beta}y^{J}.
\end{equation}

Even though reproducing well known results, our approach provides thus
a new way to look at the interaction of bosonic strings with these
topological background fields (this does not include the
background metric field describing the compactified geometry). 
While these couplings are usually
imposed by hand as external constraints with values to be
determined presumably through nonpertubative dynamics, here they
are rather understood to correspond to new intrinsic quantum degrees 
of freedom arising from a complete quantisation
of the strings, when proper account is given of the freedom 
in choosing representations for the abstract algebraic structures
defining a quantised dynamics. From that point of view, this result
also suggests that values for such topological background fields
are not to be determined through the dynamics.

Our discussion could be extended in many ways. Here it was presented
in the simplest of possible contexts, namely that of open and closed
oriented bosonic strings compactified on torii. The generalization
to superstrings may appear to be obvious at least for the
bosonic zero mode Heisenberg algebra, but a closer look
at the issue of fermionic zero modes with regards to the situation
in the bosonic sector and manifest world-sheet supersymmetry would seem
warranted nevertheless. More general settings than the simple example
discussed here, involving different interacting string sectors,
could also be investigated thoroughly, along the lines mentioned above.
It is to be expected that more general solutions 
to the conditions in (\ref{eq:Condi1}) would translate, in the T-dual 
D-brane picture, to constructions based on orbifold procedures.
And finally, extensions to other classes of non-flat compactifications
of greater relevance to string phenomenology would also deserve a
detailed study, including the case of non-oriented strings.

In parallel to such issues, it could also be of interest
to investigate, in the case of open strings, how these
quantum degrees of freedom associated to zero modes of
non-trivial topology, would translate into quantum degrees of
freedom for the closed string channel of one-loop open
string amplitudes, and {\it vice versa}. A preliminary study\cite{Payen1} of
that issue has shown that the quantum degrees of freedom discussed
in this note for both the open and the closed string sectors
are not in direct relationship through this open-closed string
duality.

\section*{\bf Acknowledgements}

We wish to thank Clifford Johnson for useful discussions
concerning D-branes and open strings. J.~G. gratefully
acknowledges Profs. Hendrik Geyer, Bernard Lategan and Frederik Scholtz
for their wonderful hospitality at the Stellenbosch Institute
for Advanced Study (STIAS) and the University of Stellenbosch (Republic
of South Africa) and their invitation to the STIAS hosted Workshop on
String Theory and Quantum Gravity (February 4--22, 2002)
where this work was initiated. This research
is partially supported by the Belgian Federal Office for Scientific,
Technical and Cultural Affairs through the Interuniversity Attraction
Pole (IAP) P5/27. J.~G. acknowledges the Abdus Salam International 
Centre for Theoretical Physics (ICTP, Trieste, Italy) Visiting Scholar 
Programme in support of a Visiting Professorship at the International 
Chair in Mathematical Physics and Applications (ICMPA, Cotonou, 
Republic of Benin). F.~P. acknowledges the support 
of the National Fund for  Scientific Research (F.N.R.S., Belgium)
through an Aspirant Research Fellowship.

\end{document}